# Application of the $O(N)$-hyperspherical harmonics to the study of the continuum limits of one-dimensional $\sigma$-models and to the generation of high-temperature expansions in higher dimensions

M. Campostrini[a], A. Cucchieri[b], T. Mendes[b], A. Pelissetto[a], P. Rossi[a], A. D. Sokal[b] and E. Vicari[a]

[a]Dipartimento di Fisica, Università di Pisa, and INFN – Sezione di Pisa, I-56100 Pisa, Italy

[b]Department of Physics, New York University, 4 Washington Place, New York, NY 10003, USA

In this talk we present the exact solution of the most general one-dimensional $O(N)$-invariant spin model taking values in the sphere $S^{N-1}$, with nearest-neighbour interactions, and we discuss the possible continuum limits. All these results are obtained using a high-temperature expansion in terms of hyperspherical harmonics. Applications in higher dimensions of the same technique are then discussed.

We will report here some new results concerning the study of one-dimensional $\sigma$-models [1] and the generation of high-temperature series in higher dimensions [2].

Both studies rely on the use of hyperspherical harmonics, for which some new formulae have been obtained [1]. Let us first review the definition of the hyperspherical harmonics. Define in $\mathbb{R}^N$ the angular momentum

$$L^{\alpha\beta} = i\left(x^\alpha \partial_\beta - x^\beta \partial_\alpha\right) \tag{1}$$

and

$$L^2 = \sum_{\alpha < \beta} L^{\alpha\beta} L^{\alpha\beta} . \tag{2}$$

The hyperspherical harmonics are simply the eigenfunctions of $L^2$ on the unit sphere $S^{N-1} \subset \mathbb{R}^N$:

$$L^2 Y_{lm}(\sigma) = \lambda_l Y_{lm}(\sigma) \tag{3}$$

with $\lambda_l = l(l + N - 2)$, $\sigma \in S^{N-1}$. The index $l = 0, 1, \ldots$ labels the eigenspaces $E_l$, and we will refer to it as *spin*. Each eigenvalue is highly degenerate: the index $m$ labels an orthonormal basis within each $E_l$. The choice of this basis is completely arbitrary but for our purposes we have found useful to use an *overcomplete* set in terms of Cartesian multipoles, i.e.

$$Y_{1,\alpha} = \sqrt{N}\,\sigma^\alpha \tag{4}$$

$$Y_{2,\alpha\beta} = \sqrt{N(N+2)/2}\left(\sigma^\alpha\sigma^\beta - \frac{1}{N}\delta^{\alpha\beta}\right) \tag{5}$$

$$Y_{l,\alpha_1\cdots\alpha_l} = \mu_l\left(\sigma^{\alpha_1}\ldots\sigma^{\alpha_l} - \text{Traces}\right) \tag{6}$$

where "Traces" is such to make $Y_{l,\alpha_1\cdots\alpha_l}$ completely symmetric and traceless and

$$\mu_l = \left[\frac{2^l\,\Gamma(l + N/2)}{l!\,\Gamma(N/2)}\right]^{1/2} \tag{7}$$

is a normalization factor. Using this representation we have been able to obtain a general expression for the Clebsch-Gordan coefficients appearing in *symmetric* tensor products of irreducible representations and scalars built from them. If we define

$$C^{l_1 l_2 l_3}_{m_1 m_2 m_3} = \int d\Omega(\sigma) Y_{l_1 m_1}(\sigma) Y_{l_2 m_2}(\sigma) Y_{l_3 m_3}(\sigma) \tag{8}$$

we have computed exact expressions for

$$C^2_{l_1 l_2 l_3} = \sum_{m_1, m_2, m_3} C^{l_1 l_2 l_3}_{m_1 m_2 m_3} C^{l_1 l_2 l_3}_{m_1 m_2 m_3} \tag{9}$$

and for the 6-$j$ symbols $\mathcal{R}(l_1, l_2, l_3; l_4, l_5, l_6)$ when one of the spins $l_i$ is 1 or 2.

Using the hyperspherical harmonics one can study the possible continuum limits of a generic one-dimensional $\sigma$-model [1]. Consider the most general Hamiltonian with nearest-neighbour couplings:

$$H = \sum_x \mathcal{V}(\sigma_x \cdot \sigma_{x+1}) \tag{10}$$

with $\sigma_x \in S^{N-1}$. The starting point is the expansion of the Boltzmann weight

$$e^{-\beta \mathcal{V}(\sigma_x \cdot \sigma_{x+1})} = F_0(\beta) \times$$
$$\left( 1 + \sum_{l=1}^{\infty} v_l(\beta) \, Y_l(\sigma_x) \cdot Y_l(\sigma_{x+1}) \right) \quad (11)$$

where $v_l(\beta)$ are coefficients that depend on the explicit form of $\mathcal{V}$ and satisfy $|v_l(\beta)| < 1$. In terms of these quantities it is easy to compute the two-point function in the spin-$k$ channel. We get

$$\langle Y_k(\sigma_0) \cdot Y_k(\sigma_x) \rangle = \mathcal{N}_k v_k(\beta)^{|x|} \quad (12)$$

where $\mathcal{N}_k$ is the dimension of $E_k$. If $0 < v_k(\beta) < 1$, the two-point function is a pure exponential and we define a mass $m_k(\beta)$ as

$$m_k(\beta) = -\log v_k(\beta) \ . \quad (13)$$

To study the critical limit we must investigate the limit $\beta \to \infty$ (no critical point can exist for finite $\beta$ in one dimension). Thus the problem of studying the possible continuum limits is reduced to the determination of the asymptotic behaviour of $v_k(\beta)$ for $\beta \to \infty$. If in this limit $m_k \to C \neq 0$ the corresponding continuum two-point function is trivial (white noise):

$$\langle Y_k(\sigma_0) \cdot Y_k(\sigma_x) \rangle_{cont} = \text{const} \times \delta(x) \ . \quad (14)$$

Thus a non-trivial continuum limit is obtained only if $m_k(\beta) \to 0$ (i.e. $v_k(\beta) \to 1$) as $\beta \to \infty$ at least for some values of $k$. The various universality classes are then characterized by the limiting mass ratios $m_k/m_l$.

A detailed study [1] has shown the existence of infinitely many families of universality classes. The *standard N-vector* universality class is obtained, for instance, by considering functions $\mathcal{V}(t)$ which have a unique absolute minimum at $t = 1$. In this case one can show that

$$m_k(\beta) = \Lambda(\beta)\lambda_k \quad (15)$$

with $\Lambda(\beta) \to 0$ for $\beta \to \infty$, so that in the continuum limit we have

$$\frac{m_k}{m_l} = \frac{\lambda_k}{\lambda_l} \ . \quad (16)$$

Analogously it is easy to recover the $RP^{N-1}$ universality class: if, for instance, $\mathcal{V}(t)$ has absolute minima at $t = \pm 1$ with $\mathcal{V}'(\pm 1) \neq 0$, then we can prove that, in the limit $\beta \to \infty$,

$$\frac{m_{2k}}{m_{2l}} = \frac{\lambda_{2k}}{\lambda_{2l}} \quad (17)$$
$$\frac{m_{2k+1}}{m_{2l}} = \infty \quad (18)$$

Besides these two cases there are many other possibilities. If, for instance, $\mathcal{V}(t)$ has minima at $t = t_0 \neq \pm 1$ and $t = 1$ and some technical conditions are satisfied [1] (physically they correspond to requiring that the contributions of the two minima to the free energy are comparable in the continuum limit), we get new universality classes in which the mass ratios are *not* related to the eigenvalues $\lambda_k$ of $L^2$.

All these results can be interpreted in another framework. In one dimension a continuum field theory is simply a continuous-time Markov process on the target manifold. Now, the generator of a continuous-time Markov process is the convex combination of a diffusion part (a second-order elliptic operator) and a jump part (a positive kernel) [3]. Physically, this means that the particle diffuses for a while according to the specified differential operator, and then, at exponentially distributed random times, jumps according to the specified probability kernel. On the sphere $S^{N-1}$ for $N \geq 3$, the only $SO(N)$-invariant second-order operator is the Laplace-Beltrami operator $L^2$ (and multiples thereof). The Markov process corresponding to pure diffusion on $S^{N-1}$ is the standard $N$-vector universality class. One can consider also processes with a jump part. It is easy to see that there is an *infinite-dimensional* family of possible $SO(N)$-invariant jump kernels $K$: indeed, one can specify an arbitrary probability distribution of jump angles $\theta \in [0, \pi]$. Correspondingly one finds an *infinite-dimensional* family of possible continuum limits. For instance the case we were mentioning above [$\mathcal{V}(t)$ with minima at $t = t_0$ and $t = 1$] corresponds to a process with diffusion and a jump distribution $\delta(\theta - \theta_0)$, $\cos \theta_0 = t_0$. Analogous considerations can be applied to $RP^{N-1}$. The *standard* $RP^{N-1}$ universality class corresponds to pure diffusion on the manifold $RP^{N-1}$, while processes with diffusion and jumps give rise to new universality classes.

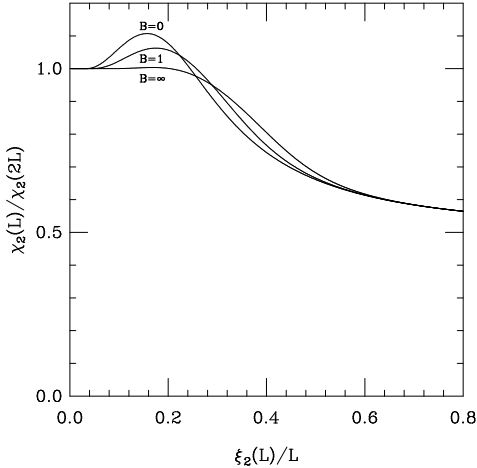

Figure 1. Finite-size-scaling function for the spin-2 susceptibility for $N = 4$. The different curves correspond to the $N$-vector universality class ($B = 0$), to the $RP^{N-1}$ universality class ($B = \infty$) while $B = 1$ corresponds to one of the intermediate universality classes introduced in [4].

In two dimensions we argued some time ago [4] that there exists a one-parameter family of universality classes for $O(N)$-invariant $\sigma$-models taking values in $S^{N-1}$, which can be obtained by taking appropriate limits in a mixed isovector/isotensor model. The same universality classes are identified in one dimension: besides being obtained as limits of a mixed model, they also arise as limits of a Hamiltonian of type (10) for suitable $\mathcal{V}$. They correspond to Markov processes on $S^{N-1}$ with diffusion and jumps of $\pi$.

For all the Hamiltonians (10), we have moreover computed [1] the finite-size-scaling (FSS) functions for the susceptibility and the correlation length in periodic boundary conditions, as well as the first correction to them. Interestingly enough, the qualitative shape of the one-dimensional FSS functions is very similar to that of their two-dimensional counterpart [4]: see Figure 1 for an example.

Let us now turn to the second subject of this talk: the problem of generating high-temperature series in dimension $d \geq 2$. We will use here the general method explained in [5]. Consider a theory with fields $\phi$ and Hamiltonian

$$H = \sum_x g(\phi_x; \beta) + \sum_{\langle xy \rangle} h(\phi_x, \phi_y; \beta) \qquad (19)$$

where the second sum is extended over all lattice links $\langle xy \rangle$, and assume a high-temperature ("strong-coupling") expansion of the form

$$e^{h(\phi_x, \phi_y; \beta)} = F_0(\beta) \left[1 + \sum_n h_n(\phi_x, \phi_y; \beta)\right] \qquad (20)$$

with $h_n(\phi_x, \phi_y; \beta) \sim \beta^n$ for $\beta \to 0$. In the present implementation of the program one must assume an additional property: that the integral

$$\int d\phi_x \, e^{g(\phi_x; \beta)} h_{n_1}(\phi_x, \phi_{y_1}; \beta) \cdots h_{n_k}(\phi_x, \phi_{y_k}; \beta) (21)$$

vanishes if $\sum_i n_i$ is odd. This property allows us to relate every non-vanishing high-temperature graph to one or more lattice random walks (RW). However, we must notice that there are interesting models for which this property does not hold: for instance, the $RP^{N-1}$ and the $CP^{N-1}$ $\sigma$-models. An additional simplification can be obtained if one requires an orthogonality property:

$$\int d\phi_x \, e^{g(\phi_x; \beta)} h_{n_1}(\phi_x, \phi_{y_1}; \beta) h_{n_2}(\phi_x, \phi_{y_2}; \beta) = 0 (22)$$

if $n_1 \neq n_2$. In this case all high-temperature graphs can be generated by considering only non-reversal RW. We refer to [5] for details on the practical implementation.

For this very general class of theories we have computed [2] the two-point functions

$$G_1(x) = \langle h_1(\phi_0, \phi_x; \beta) \rangle \qquad (23)$$

$$G_2(x) = \langle h_2(\phi_0, \phi_x; \beta) \rangle \qquad (24)$$

for various lattices in $d = 2, 3$. The result is written as a sum of abstract graphs which have a pair of integers $(n, L)$ associated to each link (in [5] these diagrams were called *skeleta*): the integer $n$ indicates the spin associated to the link, while $L$ is its length. For instance, to the graph appearing in Figure 2 we associate the expression

$$\prod_{k=0}^{L_1-1} h_{n_1}(\phi_k^1, \phi_{k+1}^1; \beta) \prod_{h=0}^{L_2-1} h_{n_2}(\phi_h^2, \phi_{h+1}^2; \beta) \times$$

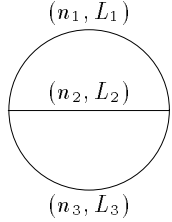

Figure 2. A simple strong-coupling graph.

$$\prod_{l=0}^{L_3-1} h_{n_3}(\phi_l^3, \phi_{l+1}^3; \beta) \qquad (25)$$

with $\phi_0^1 = \phi_0^2 = \phi_0^3$ and $\phi_{L_1}^1 = \phi_{L_2}^2 = \phi_{L_3}^3$. This expression must then be integrated over all fields with measure $d\phi \exp[g(\phi; \beta)]$.

To go further one must specify the fields and the specific model and compute the explicit contribution of each skeleton. For $O(N)$-invariant $\sigma$-models with nearest-neighbour interactions we write

$$H = \sum_{\langle xy \rangle} \mathcal{V}(\sigma_x \cdot \sigma_y; \beta) \qquad (26)$$

and expand $\mathcal{V}(\sigma_x \cdot \sigma_y; \beta)$ as in (11); then we express all Green's functions in terms of the *pseudocharacter* coefficients $v_l(\beta)$. What must then be computed is the group-theoretic factor of each skeleton. We have written a program in MATH-EMATICA that does this for an arbitrary skeleton. If the spin associated to each link is small ($l \le 4$), the program is very efficient: the most complicated skeleta require $\approx 1$ min of CPU-time on a Sun4/330 with 24 Mbytes of RAM.

The computation can be simplified if one uses the explicit knowledge of $C^2_{l_1 l_2 l_3}$ and of the 6-$j$ symbols $\mathcal{R}(l_1, l_2, l_3; l_4, l_5, l_6)$. Indeed, one can apply recursively the rules appearing in Figure 3 where

$$A_{ij;kl} = \frac{C^2_{ikl}}{\mathcal{N}_i} \delta_{ij} \qquad (27)$$

$$B_{ijk;lmn} = \frac{\mathcal{R}(j, i, k; l, m, n)}{C^2_{ijk}} \qquad (28)$$

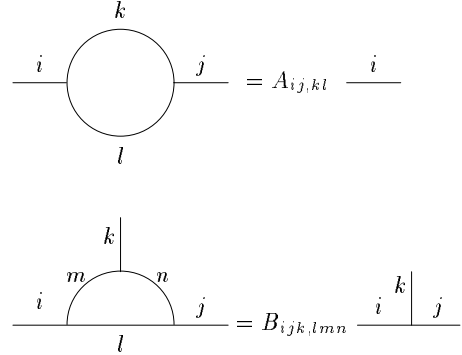

Figure 3. Reduction relations for bubbles and triangles.

These two rules allow the complete computation of nearly all the graphs which appear in our expansions; only for a few of the graphs higher-order symbols need to be computed.

This work was supported in part by INFN and NSF grant DMS-9200719.